\documentclass[conference]{IEEEtran}
\IEEEoverridecommandlockouts

\usepackage{cite}
\usepackage{amsmath,amssymb,amsfonts}
\usepackage{bm}
\usepackage{algorithmic}
\usepackage{graphicx}
\usepackage{textcomp}
\usepackage{xcolor}
\usepackage{url}
\usepackage[caption=false]{subfig}
\usepackage{pbalance}
\usepackage{cleveref}
\def\BibTeX{{\rm B\kern-.05em{\sc i\kern-.025em b}\kern-.08em
    T\kern-.1667em\lower.7ex\hbox{E}\kern-.125emX}}
\begin{document}

\title{Comparing User Activity on X and Mastodon%
\thanks{The research was supported by ROIS NII Open Collaborative Research 2024-24FS02.}%
}

\author{\IEEEauthorblockN{Shiori Hironaka}
\IEEEauthorblockA{%
\textit{Kyoto University}\\
Kyoto, Japan \\
hironaka@media.kyoto-u.ac.jp\\
0000-0001-7994-2858}
\and
\IEEEauthorblockN{Mitsuo Yoshida}
\IEEEauthorblockA{%
\textit{University of Tsukuba}\\
Tokyo, Japan \\
mitsuo@gssm.otsuka.tsukuba.ac.jp\\
0000-0002-0735-1116}
\and
\IEEEauthorblockN{Kazuyuki Shudo}
\IEEEauthorblockA{%
\textit{Kyoto University}\\
Kyoto, Japan \\
0000-0002-3939-9800}
}
\maketitle

\begin{abstract}
The ``Fediverse'', a federation of decentralized social media servers, has emerged after a decade in which centralized platforms like X (formerly Twitter) have dominated the landscape. The structure of a federation should affect user activity, as a user selects a server to access the Fediverse and posts are distributed along the structure. This paper reports on the differences in user activity between Twitter and Mastodon, a prominent example of decentralized social media. The target of the analysis is Japanese posts because both Twitter and Mastodon are actively used especially in Japan. Our findings include a larger number of replies on Twitter, more consistent user engagement on mstdn.jp, and different topic preferences on each server.
\end{abstract}


\section{Introduction}

Social media has become an integral part of people's daily lives, serving as a platform for communication, information gathering, and content sharing.
These platforms generate vast amount of data that capture various aspects of human activity and interaction.
Researchers have utilized social media data to analyze social trends~\cite{Benhardus2013,Bogdanowicz2022} and gain insights into public opinions and reactions to social issues~\cite{Kushin2019,ChouJen2020,Hu2023}.

Traditionally, centralized social media platforms like X\footnote{\url{https://x.com/} (accessed 2024-10-15)} (formerly Twitter) have dominated the landscape, with a single company acting as the service provider.
However, recent years have seen the emergence and growth of decentralized social media platforms.
These decentralized networks, often referred to as the ``Fediverse,'' consist of multiple servers operated by different service providers.
They form a large, interconnected social network using a unified protocol implemented across server clusters.
Users can choose and connect to their preferred servers to access these services.

One prominent example of decentralized social media is Mastodon\footnote{\url{https://joinmastodon.org/} (accessed 2024-10-15)}, which implements the ActivityPub protocol.
While the user base of decentralized platforms has been growing, it remains unclear whether their user activities mirror those of established centralized platforms like Twitter.

Japan presents an interesting case study in this context. As Twitter's second-largest market after the United States, it has been widely adopted for various purposes, including accessing real-time news and updates, communicating with friends, and sharing opinions on current events and hobbies. Concurrently, Japan has experienced significant adoption of decentralized social media platforms.

This research aims to investigate the differences in user activities between Twitter and Mastodon from a user and topic perspective.
By examining these platforms, we seek to understand how user activities differ between centralised and decentralised social media ecosystems.
Our findings include a larger number of replies on Twitter, more consistent user engagement on mstdn.jp, and different topic preferences on each server.

\section{Data}

\subsection{Collection}

We collect data using streaming API of Twitter and Mastodon.
Mastodon is a decentralized social media platform where services are independent of each other in units called instances (servers).
Users are free to create and operate instances.
Even if users do not open an instance, users can use social media by joining other people's instances.
When collecting data on Mastodon, it is necessary to collect data for each instance separately.
In this paper, we used data from mstdn.jp, the largest Mastodon instance in Japan, and pawoo.net an instance that focuses on topics related to illustration.

From Twitter, posts are collected using the 1\% sampled streaming API, limited to Japanese.
Local timelines were collected from two instances of Mastodon, mstdn.jp and pawoo.net.
An instance's local timeline contains posts from users within that instance whose posts are set to public.
Although Mastodon does not limit the language of posts to Japanese.%

As a result, we collected 125,703,990 posts from Twitter, 999,392 posts from mstdn.jp, and 425,138 posts from pawoo.net, between May 3 and June 23, 2023\footnote{After June 23, 2023, Twitter's API is no longer available.}.

\subsection{Aggregation}

The data we collected consists of streams of posts.
Each post is accompanied by the user's profile information at the time of posting.
Therefore, we aggregate these posts by user and extract the most recent profile information for each user.

There are three types of posts:
\begin{itemize}
    \item Original posts: Posts that are neither replies nor shares.
    \item Reply posts: Posts with a reply-to post ID in their metadata.
    \item Share posts: Posts referred to as ``boosts'' on Mastodon and ``reposts'' or ``retweets'' on Twitter.
\end{itemize}

For each user, we calculate the following user attributes:
\begin{itemize}
\item Total posts during the period: The number of posts made by the user that appeared in the stream during the collection period.
\item Proportion of reply posts: The ratio of reply posts to the total number of collected posts for the user.
\item Active days: The number of days between the user's first and last posts within the collection period.
\end{itemize}
We conduct analyses for each user attributes.
In addition to these attributes, we conduct analyses using the number of followees and followers, and follower--followee ratio.

\subsection{Limitation}

The data used in this analysis was collected through streaming APIs. There are several limitations to consider:
\begin{itemize}
    \item For Twitter, we used a 1\% sampled data stream, resulting in a lower proportion of collected posts compared to Mastodon.
    
    \item Mastodon data was collected from local timelines. Consequently, reblogged posts are not included in our dataset.
    
    \item While the Twitter data consists entirely of Japanese posts due to language filtering during collection, the Mastodon data was collected without language restrictions. As a result, the Mastodon dataset may include posts in languages other than Japanese.
\end{itemize}

These limitations should be taken into account when interpreting the results of our comparative analysis between Twitter and Mastodon.

\section{Analysis of User Attributes}

\subsection{Total Posts During the Period}

\begin{figure*}[tp]
    \centering
    \includegraphics[width=\linewidth]{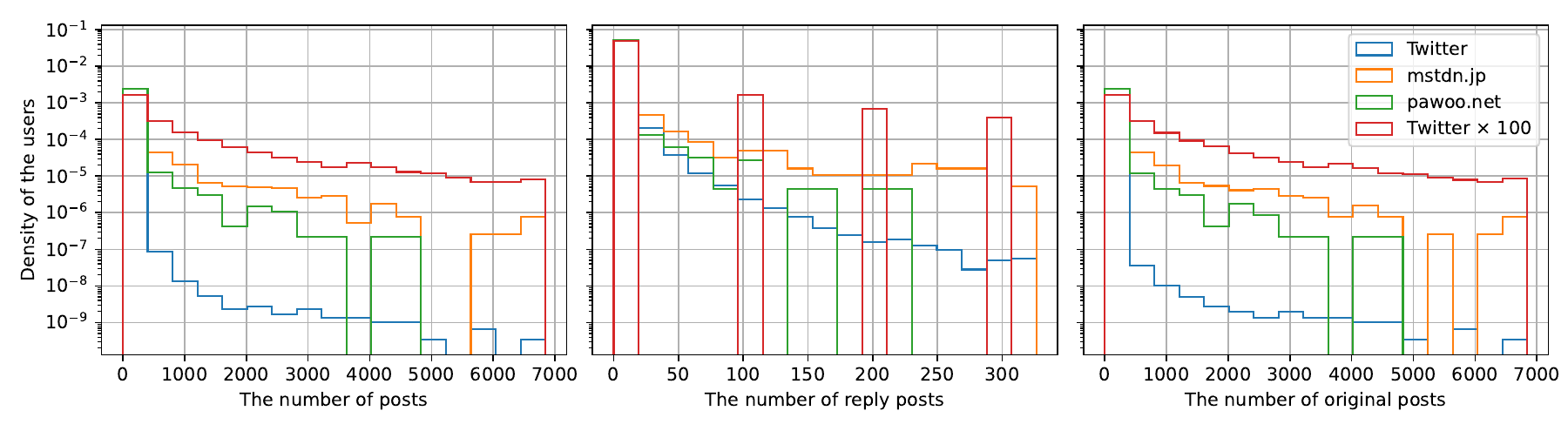}%
    \caption{Density histogram of the number of posts, reply posts, and original posts.}%
    \label{fig:dist-posts}
\end{figure*}

Platforms with a high number of posts are considered actively used platforms.
First, \Cref{fig:dist-posts} shows the probability density histograms of the number of posts for each post type.
Share posts do not exist on Mastodon, so they are not used in this study.
The number of posts refers to the total number of posts across all post types.

From \Cref{fig:dist-posts}, it can be observed that Twitter tends to have many users with a low number of posts.
Only Twitter data is 1\% sampled, while Mastodon data is 100\% complete.
As a result, many Twitter users have no observed posts, and even when posts are observed, only $1/100$ of the actual posts are captured. This is likely the cause of the observed distribution.

Since Twitter posts may actually exist at 100 times the observed rate, \Cref{fig:dist-posts} also includes a plot where the observed number of Twitter posts is multiplied by 100.
Considering any type of post, it can be inferred that Twitter has the highest number of posts.
Comparing Mastodon instances, it was found that mstdn.jp tends to have more posts than pawoo.net.

\subsection{Proportion of Reply Posts}

Replies constitute a form of direct communication in social media, facilitating user-to-user conversations.
A high proportion of replies indicates that a platform is used for direct interactions and dialogic communication.

\begin{figure}[tp]
    \centering
    \includegraphics[width=\linewidth]{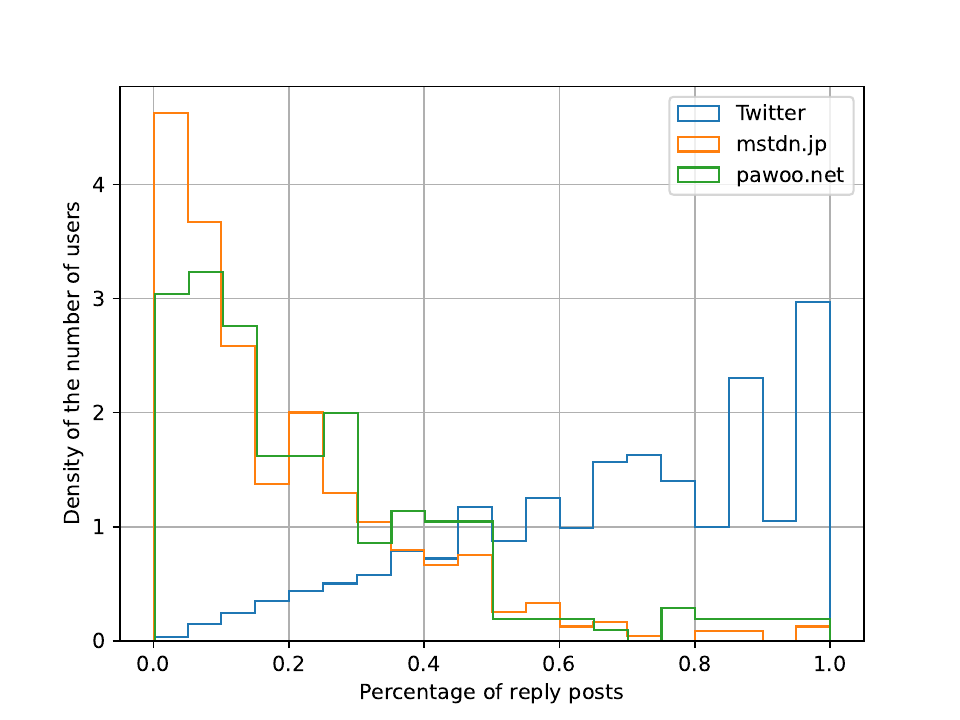}%
    \caption{Density histogram of the percentage of reply posts across platforms. Twitter tends to have a higher proportion of reply posts compared to Mastodon.}%
    \label{fig:dist-reply}
\end{figure}

We calculated the ratio of reply posts to total posts for each user in our collected data.
Since only a small fraction of users utilize replies, and displaying data for all users would make the results difficult to interpret, we focused on users who actively use replies.
We plotted data only for users with more than 5 reply posts, as shown in \Cref{fig:dist-reply}.
Among users who posted at least once during the collection period, the percentage of users with more than 5 reply posts was 4.1\% for Twitter, 5.0\% for mstdn.jp, and 1.8\% for pawoo.net.
These users are represented in the plot.

\Cref{fig:dist-reply} reveals that Twitter tends to have a higher proportion of reply posts compared to Mastodon.
This result may be attributed to either more frequent use of replies on Twitter or fewer replies appearing on Mastodon's public timeline.
Mastodon offers more flexible privacy settings for posts compared to Twitter.
Consequently, Mastodon users may choose not to display their replies on the local timeline, potentially resulting in fewer replies in our aggregated data.
Twitter lacks such granular settings; thus, if an account is public, replies are displayed on the public timeline by default.

\subsection{Active Days}

To examine whether platforms are used continuously, we calculated the active days for each user.
We define active days as the number of days between a user's first and last observed posts in our collected data.
Users with fewer than two posts are assigned zero active days.
Given our total data collection period of 52 days, the maximum possible active days is 52.

\begin{figure}[tp]
    \centering
    \includegraphics[width=\linewidth]{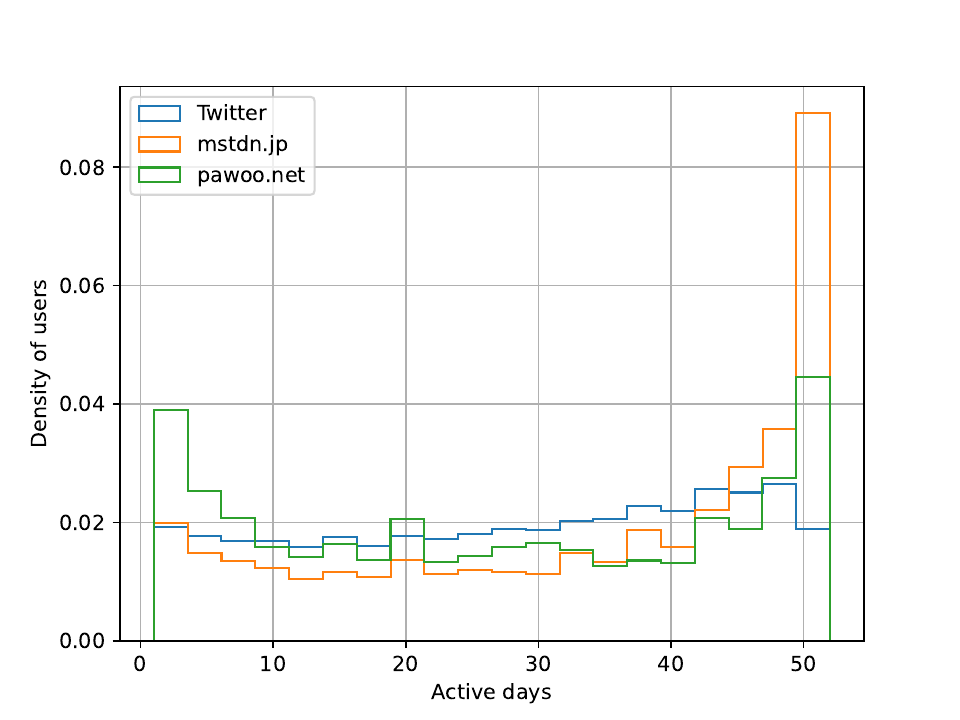}%
    \caption{Density histogram of active days across platforms. Users on mstdn.jp exhibit a tendency towards more regular platform engagement.}%
    \label{fig:dist-active-days}
\end{figure}

\Cref{fig:dist-active-days} presents the probability density histogram of active days.
The results indicate that mstdn.jp exhibits a higher proportion of users with longer active days, suggesting more consistent and enthusiastic user engagement compared to other servers.
Conversely, pawoo.net demonstrates a higher proportion of users with active days of just a few days.
Twitter falls between these two extremes; however, direct comparison is not feasible due to the 1\% sampling of Twitter data.

\subsection{Degree Distribution}

\begin{figure}
    \centering
    \subfloat[Number of followees.]{%
        \includegraphics[width=\linewidth]{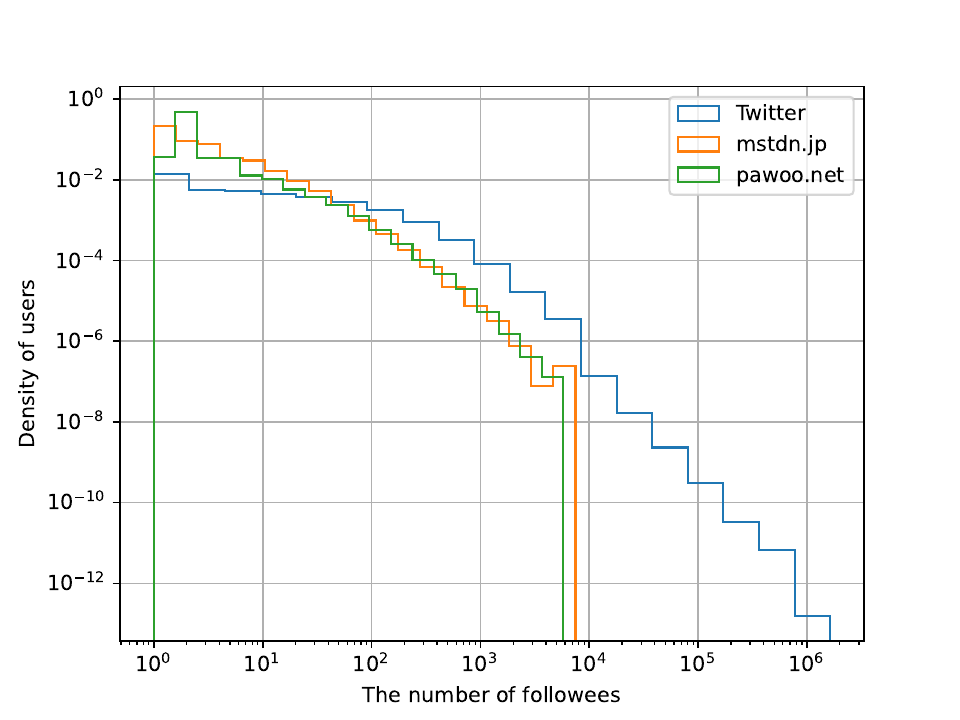}%
        \label{fig:dist-friends}%
    }\\
    \subfloat[Number of followers.]{%
        \includegraphics[width=\linewidth]{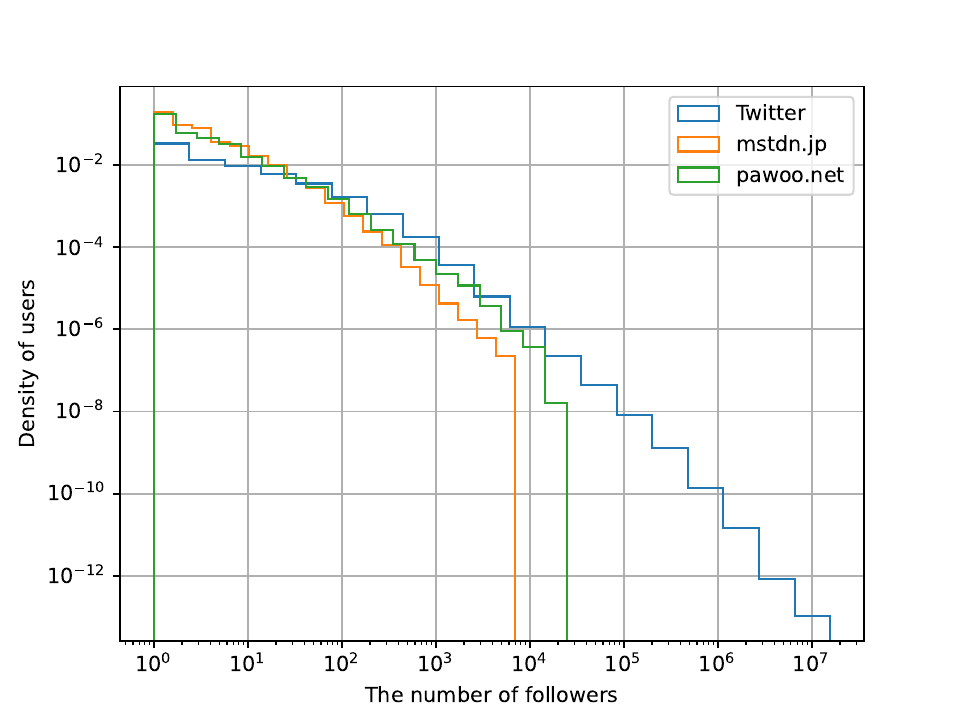}%
        \label{fig:dist-followers}%
    }
    \caption{Density histogram of number of followees and followers across platforms.}
    \label{fig:dist-follow}
\end{figure}

We extracted the number of followees and followers for each user from their profile information.
Due to the difference in social network sizes between Twitter and the Fediverse, the scale of followees and followers for each user also differs.
To compare the distribution of followees and followers across platforms, we plotted probability density histograms.

The results are shown in \Cref{fig:dist-follow}.
Our analysis revealed that mstdn.jp exhibited the highest proportion of users with zero followees, while pawoo.net demonstrated the highest proportion of users with zero followers.
Furthermore, mstdn.jp tended to have fewer users with a high number of followers.
Moreover, both Mastodon instances (mstdn.jp and pawoo.net) showed a tendency towards having more users with fewer followees and followers compared to Twitter. This observation may indicate that the Fediverse is still an emerging social network that has not yet reached its full potential in terms of user connections and network density.

\subsection{Follower--Followee Ratio}

The number of followees and followers are key characteristics of a user's established social network, reflecting how users utilize social media platforms. However, these values are influenced by the total number of users on each platform, making direct comparisons between Twitter and the Fediverse challenging.

Previous studies have classified users based on their follower--followee ratio~\cite{Yan2018,Oshimo2022}.
Therefore, we calculate the ratio of followees to followers.
The follower--followee ratio is calculated using the following formula:
\begin{equation}
    \mathrm{Ratio} = \log_{10}\left(\frac{\text{Number of Followees} + 1}{\text{Number of Followers} + 1}\right)
\end{equation}
We apply a logarithmic transformation to the ratio. Consequently, users with positive values have more followees than followers, while users with negative values have more followers than followees.

\begin{figure}[tp]
    \centering
    \includegraphics[width=\linewidth]{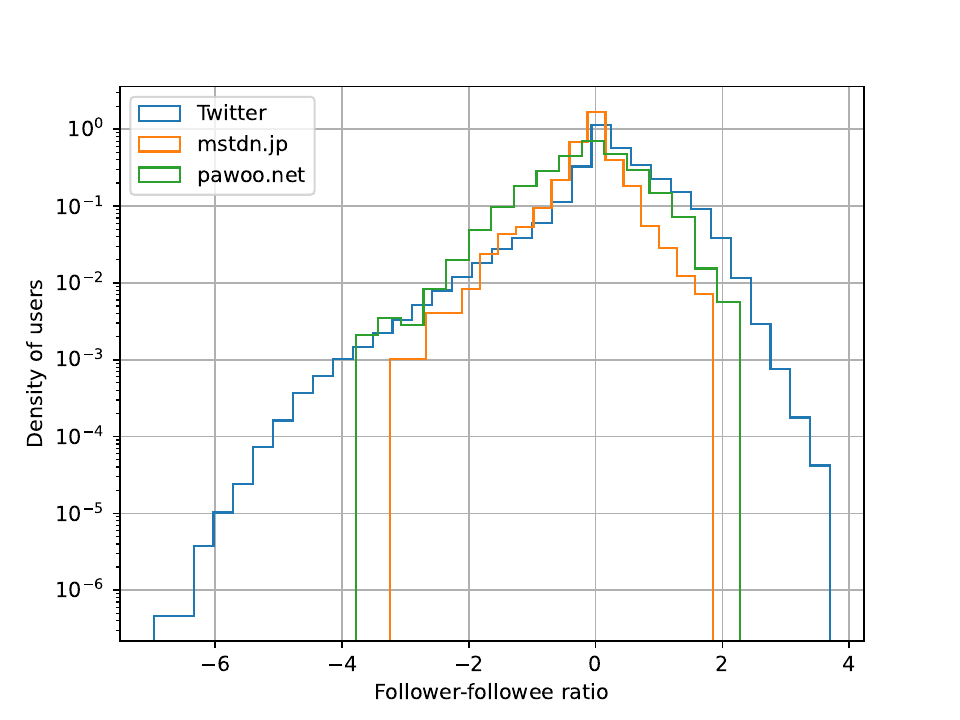}%
    \caption{Density histogram of follower--followee ratios across platforms.}%
    \label{fig:dist-follow-ratio}
\end{figure}

\Cref{fig:dist-follow-ratio} shows the probability density histogram of the calculated follower--followee ratios for each platform.
To exclude inactive users, we only plotted data for users with a combined total of followees and followers greater than 20.
Our analysis reveals several key observations. First, mstdn.jp exhibits a high proportion of users with follower--followee ratios close to 0, indicating a good balance between followees and followers. This suggests that these users primarily use social media for interaction purposes.
Second, Twitter has more users with positive values compared to other platforms, indicating a higher proportion of users with subscribing purposes. Lastly, pawoo.net has more users with negative values, suggesting the presence of some influential users on this platform.
These findings highlight the different user behaviors and network structures across the studied platforms, reflecting diverse user activities and social dynamics.
The variations in follower--followee ratios provide insights into how users on each platform engage with content and build their social networks.

\section{Analysis of Topics}

To compare across platforms, we first train topic models.
Then, using the estimated topics, we investigate the characteristics of topics in posted content for each platform.

\subsection{Data Preprocess}

To prepare the collected posts for analysis, the following preprocessing steps were performed:
\begin{enumerate}
\item Extract post content and normalize strings.
\item Determine vocabulary.
\item Create a corpus for topic analysis.
\end{enumerate}

First, reposts and boosts (both features similar to retweets) were excluded from the collected posts.
Since Mastodon posts contain HTML tags, these tags were removed.
Additionally, URLs and user IDs were removed from both Twitter and Mastodon posts, and string normalization was performed using neologdn\footnote{\url{https://github.com/ikegami-yukino/neologdn}}.

Next, Japanese morphological analysis was then performed using Vibrato\footnote{\url{https://github.com/daac-tools/vibrato}} and mecab-ipadic v2.7.0.
Words whose parts of speech were identified as nouns other than non-independent, verbs other than non-independent or suffixes, adjectives, adverbs, adnominals, interjections, or symbols were kept in the documents, while others were removed.
Verbs were lemmatized to their base forms.
Single-character hiragana or alphabet words and words composed solely of numbers were also removed.
Words appearing in more than 10\% of all posts were excluded, and the top 50,000 words with the highest document frequency were selected as the vocabulary.

A corpus of documents was created that contained only words from the vocabulary present in each post.
Posts that did not contain words from the vocabulary were excluded.

Finally, we got 17,432,493 posts from Twitter, 767,746 posts from mstdn.jp, and 268,614 posts from pawoo.net.
The number of posts from Twitter is more than ten times higher than that from Mastodon.
The average number of words per document after preprocessing was 8.01 for Twitter, 8.81 for mstdn.jp, and 10.55 for pawoo.net.

\subsection{Preparation of Topic Model}

The Biterm Topic Model (BTM)~\cite{Cheng2014} is employed for our analysis.
BTM has been reported to be particularly effective for short texts~\cite{Qiang2022}.
Given that this study involves topic analysis of over 18 million posts, it is crucial to utilize an efficient method to handle such a large dataset.
We experimented with various numbers of topics ($K$) for the BTM. Ultimately, we determined that $K=30$ was sufficient to capture the essential characteristics of the platform.
Therefore, this number of topics was adopted for our analysis.

Suppose that given corpus $D~(d \in D)$ and the number of topics $K$, $\theta_{k}$, the probability of topic $k$, and $\phi_{kv}$, the probability of word $v$ choosing topic $k$, can be learned.
Then, from these parameters, the topic distribution $P(z=k \mid d)$ for a document can be estimated.
In this study, we used the bitermplus\footnote{\url{https://github.com/maximtrp/bitermplus}} package for BTM computations, with hyperparameters set to $\alpha = 6.25$ and $\beta = 0.01$.

To examine the characteristics of a set of documents, we define the topic distribution for the document set.
The topic distribution for document set $\bm{D}_F$ is calculated using the following equation:
\begin{equation}
  \frac{1}{N_F} \sum_{d \in \bm{D}_F} P(z=k \mid d) \label{eq:topicdist}
\end{equation}
where $N_F$ is the number of posts in $\bm{D}_F$ ($|\bm{D}_F| = N_F$).

\subsection{Result}

\begin{figure}[tp]
    \centering
    \includegraphics[width=\linewidth]{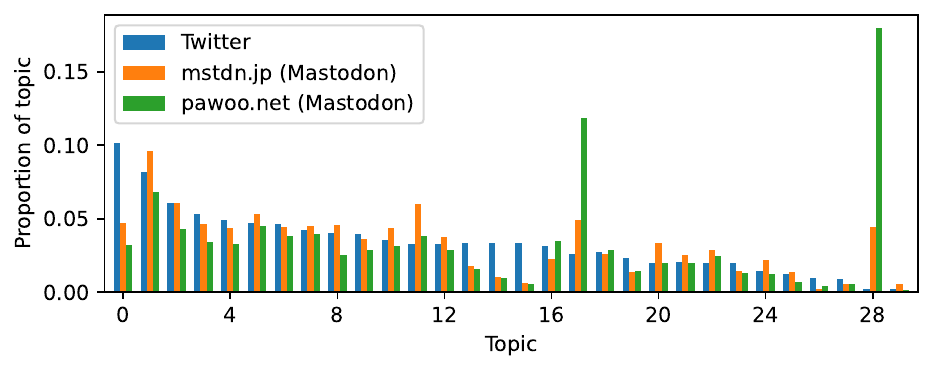}%
    \caption{Topic distributions for each platform ($K=30$).}%
    \label{fig:topicdist30}
\end{figure}

We investigate the characteristics of the posted content for each platform using traind model.
The topic distributions for each platform, computed from the BTM with $K=30$, are shown in \Cref{fig:topicdist30}.
Overall, the broad trends of posting topics for each platform have been captured.

According to \Cref{fig:topicdist30}, the most common topics in Twitter's posts are Topics 0, 13, 14, 15, and 26.
Topic 0 consists mainly of emojis and words expressing happiness and support for content creators or favorites.
Topic 13 is about Gacha games and events, Topic 14 is about exchanging goods, Topic 15 is about Seven Eleven-related advocacy campaigns, and Topic 26 is about Lawson-related advocacy campaigns.
It is noteworthy that campaigns to encourage user contributions as part of corporate marketing strategies are prevalent on Twitter, but not observed on Mastodon.

According to \Cref{fig:topicdist30}, the prevalent topics in mstdn.jp posts are Topic 1, Topic 10, and Topic 11.
Topic 1 consists of everyday posts, Topic 10 contains words related to life and family, and Topic 11 contains words related to society and politics, with mature expressions.

According to \Cref{fig:topicdist30}, the most common topics in pawoo.net posts are Topic 17 and Topic 28, with Topic 16 being as common as Twitter.
Topic 17 contains words related to anime, AI, illustration, Topic 28 contains mostly English words, and Topic 16 contains words related to the illustration community.
This reflects the preference of the Mastodon instance for creative activities centered around illustration.

\section{Related Work}

The study of user activities on social media platforms has been a long-standing area of research in the field of social computing~\cite{Poblete2011,Longley2015,Das2016,Luo2019,Valkenburg2022,Xue2023}.
These patterns provide valuable insights into user behavior, information dissemination, and community dynamics.
However, as Trifiro and Gerson~\cite{Trifiro2019} point out, many existing methodologies are constrained to single social media platforms, highlighting a significant gap in approaches that can capture cross-platform or generalized usage patterns.

In recent years, the landscape of social media has evolved with the emergence and growth of decentralized platforms.
These platforms, exemplified by Mastodon and others in the Fediverse, introduce unique characteristics that set them apart from traditional centralized platforms like Twitter~\cite{Raman2019}.
Several studies have begun to explore these decentralized ecosystems. Zignani et al.~\cite{Zignani2019} provided one of the first comprehensive analyses of the Mastodon network structure, while La Cava et al.~\cite{LaCava2021,LaCava2022} explored the specific features and user behaviors within decentralized social media environments.
Khateeb et al.~\cite{Al-khateeb2022} also analyzed trends within a single Mastodon instance.
To the best of our knowledge, no research has yet undertaken a comprehensive comparison of user activities between Twitter and Mastodon, nor attempted to compare topics across multiple Mastodon instances.
Our study aims to address this gap by providing the first direct comparison of user activities between Twitter and Mastodon.

\section{Conclusion}

This study aimed to investigate the differences in user activities between centralized (Twitter) and decentralized (Mastodon) social media platforms, focusing on user attributes and topic distributions. Our analysis revealed several key findings:
\begin{enumerate}
    \item Post frequency: Twitter users generally posted more frequently than Mastodon users, with mstdn.jp users being more active than pawoo.net users.
    \item Reply behavior: Twitter showed a higher proportion of reply posts compared to Mastodon instances, suggesting more direct user-to-user interactions.
    \item User engagement: mstdn.jp demonstrated more consistent user engagement over time compared to pawoo.net and Twitter.
    \item Network structure: Mastodon instances showed a tendency towards smaller number of followers and followees compared to Twitter, possibly reflecting the nascent stage of the Fediverse.
    \item Topic distribution: Each platform exhibited distinct topic preferences, with Twitter featuring more marketing campaigns, mstdn.jp focusing on everyday life and societal issues, and pawoo.net centering around creative activities.
\end{enumerate}

These findings highlight the unique characteristics of centralized and decentralized social media platforms, reflecting differences in user behavior, community dynamics, and content focus. Our research contributes to the understanding of how decentralized social media ecosystems differ from traditional centralized platforms, providing insights for future development and research in this evolving landscape.
Furthermore, this study provides insights into key aspects for future research to focus on when comparing different social media platforms.

\bibliographystyle{IEEEtran}
\bibliography{./reference}

\end{document}